\documentclass[twocolumn,aps,pra,superscriptaddress,showpacs]{revtex4-1}
\usepackage{epsfig}
\usepackage[flushleft]{threeparttable}
\usepackage[colorlinks=true,citecolor=blue,urlcolor=blue,linkcolor=blue]{hyperref}
\usepackage{natbib}
\usepackage[normalem]{ulem}
\usepackage[usenames,dvipsnames]{xcolor}
\usepackage{color} 
\definecolor{darkgreen}{rgb}{0.0, 0.6, 0.0}

\newcommand{\barium}{$^{138}$\textrm{Ba}$^+$}

\begin{document}

\title{Increase of barium ion-trap lifetime via photodissociation}%

\author{Hao Wu}
\affiliation{Department of Physics and Astronomy, University of California, Los Angeles, California 90095, USA}
\affiliation{Challenge Institute for Quantum Computation, University of California, Los Angeles, California 90095, USA}
\author{Michael Mills}
\affiliation{Department of Physics and Astronomy, University of California, Los Angeles, California 90095, USA}

\author{Elizabeth West}
\affiliation{Department of Physics and Astronomy, University of California, Los Angeles, California 90095, USA}

\author{Michael C. Heaven}
\affiliation{Department of Chemistry, Emory University, Atlanta, Georgia 30322, USA}

\author{Eric R. Hudson}
\affiliation{Department of Physics and Astronomy, University of California, Los Angeles, California 90095, USA}

\affiliation{Challenge Institute for Quantum Computation, University of California, Los Angeles, California 90095, USA}

\affiliation{Center for Quantum Science and Engineering, University of California, Los Angeles, California 90095, USA}
\date{\today}


\begin{abstract}

The lifetime of Ba$^+$ ions confined in a Paul trap is found, under typical conditions, to be limited by chemical reactions with residual background gas. 
An integrated ion trap and time-of-flight mass spectrometer are used to analyze the reactions of the trapped Ba$^+$ ions with three common gases in an ultrahigh vacuum system (H$_2$, CO$_2$ and H$_2$O). 
It is found that the products of these reactions can all be photodissociated by a single ultraviolet laser at 225~nm, thereby allowing the recovery of the Ba$^+$ ions and leading to an increase of the effective trap lifetime. 
For a Coulomb crystal, the lifetime increased from roughly 6~hours to 2~days at room temperature. 
It is suggested that higher enhancement factors are possible in systems with stronger traps.
In addition, photodissociation wavelengths for other common trapped ion systems are provided.


\end{abstract}

\maketitle

To date, the highest-fidelity quantum operations and  the largest ratio of qubit coherence time to gate time has been achieved in trapped ion systems~\cite{doi:10.1063/1.5088164,Christensen2020,Ballance2016,Srinivas2021,Gaebler2016}. 
As such, ion-based qubits are a leading candidate for the construction of a large-scale quantum information platform, with major efforts underway using integrated photonics and microfabricated traps~\cite{Bruzewicz2019,Mehta2020}.  
However, as these systems are scaled to a larger number of ions, the finite lifetime of ions in the trap, $\tau$, leads to probability for loss of an ion $P = 1-e^{-N t/\tau}$, where $N$ is the number of ions in the register and $t$ is the time required for the entire quantum operation.
While loss of an ion is generally detrimental to the operation at hand, it also triggers a time-consuming reloading operation.
Further, the reloading process can often cause a number of deleterious effects, including unwanted charging of the trap electrodes. 
Several techniques have been developed to mitigate these problems, including loading from a magneto-optical trap~\cite{Bruzewicz2016} and placing the ion trap in a cryogenic environment~\cite{Pagano_2018,Brandl2016,Leopold2019}, where the residual background gas, which is typically assumed to be responsible for ion loss, is greatly reduced. 

Here, we study the trap-loss process for Ba$^+$ ions and find that ion loss is dominated by chemical reactions with residual background gas which produce molecular ions that, at high-enough trapping potential and with efficient sympathetic cooling, do not leave the ion trap.
Building on the work of Ref.~\cite{Brian2015,Kuang2011} and guided by available spectroscopic data and \textit{ab initio} molecular structure calculations, we demonstrate a simple technique for recovering the atomic ion qubit from the product molecular ion.
Specifically, we employ photodissociation with a single laser that is capable of dissociating all common molecular ion products and increases the observed trapped-ion lifetime by a factor of $2.3(1)$ in an ion chain and 7.6(8) for an ion crystal\textemdash~here () denotes one standard error. The enhancement appears to be limited by the inability of the current ion trap to capture all of the recoiling product molecular ions. 
Systems with higher secular frequency~\cite{Guggemos_2015} will likely realize larger enhancement factors.


Besides reactive loss, micromotion interruption due to background gas represents another possible loss mechanism for ions.  
The time-varying nature of the ion trap confinement provides mechanisms for energy to be coupled from the radio-frequency (rf) electric potential that confines the ions into the ion motion during a collision with background gas, as well as with other ions.
Refs.~\cite{Kuang143009,Kuang173003,Bifurcation,Cetina2012} analyzed this phenomenon and found that it is possible to achieve ion temperatures several times that of the background gas through such collisions. 
Therefore, elastic collisions with the background gas and then further collisions between the ions, can lead to ion loss. 

To investigate the role of micromotion-interruption-induced loss,
$^{138}$Ba$^+$ are trapped and laser-cooled in a linear Paul trap with field radius $r_o = 6.85$~mm, driven with an rf frequency at $\Omega = 2\pi \cdot 0.7$~MHz and a peak-to-peak amplitude of 320~V, leading to a radial secular frequency of $\omega_r \approx 2\pi \cdot 70$~kHz.
Axial confinement is provided by two DC electrodes spaced by 20~mm.
The ions are detected either via imaging their laser-cooling-induced fluorescence through an objective with numerical aperture of 0.23 or by an integrated time-of-flight mass spectrometer\textemdash~further details are provided in Ref.~\cite{MikeThesis}.

\begin{figure}[t]
\includegraphics[width=\linewidth]{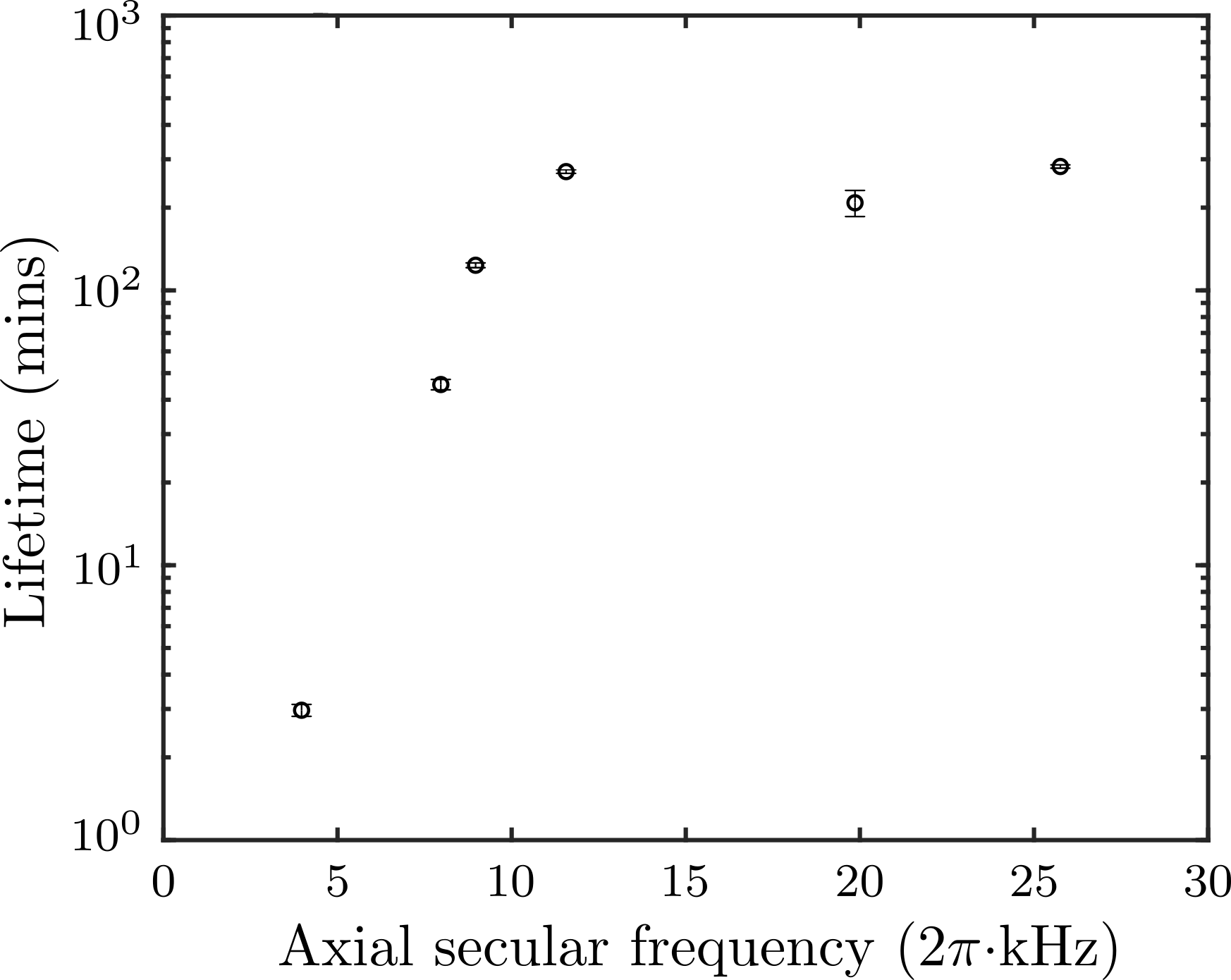}%
\vspace{-5pt}
\caption{\label{endcap}
$^{138}$Ba$^+$ trap lifetime at different axial secular frequencies, which correspond to different trap depths. The lifetime is extracted from an exponential fit.
Error bars represent one standard error extracted from an exponential fit to the observed trap decay of a single experimental trial.
\vspace{-4mm}}
\end{figure}

The lifetime of ions in the trap is shown in Fig.~\ref{endcap} for several values of the axial secular frequency of a single trapped $^{138}$Ba$^+$. The axial frequency is measured by applying an oscillating voltage to one of the endcaps, which resonantly excites the ion motion~\cite{Home_2011}. When the axial secular frequency is larger than $\omega_z =2\pi \cdot 11$~kHz, the micromotion-interruption-induced loss appears to be mostly suppressed due to the increased trap depth. 
In the subsequent experiments, the axial secular frequency is fixed to $\omega_z = 2\pi \cdot 25$~kHz.

With the role of micromotion interruption clarified, the remaining trap loss is presumed to be predominantly due to chemical reactions with the residual background gas that can, potentially, both release a large amount of energy and convert the atomic ion into a molecular compound. 
To investigate this, we expose a trapped sample of $^{138}$Ba$^+$ to one of three common UHV residual gases H$_2$, CO$_2$, and H$_2$O, and monitor the trap loss -- although CO is also a common residual gas it does not react with Ba$^+$~\cite{Roth2008} and is not studied here.
Using an integrated time-of-flight mass spectrometer (ToF-MS)~\cite{TOF}, we monitor the appearance of any ion products of the chemical reactions.

\begin{figure}[t]
\includegraphics[width=\linewidth]{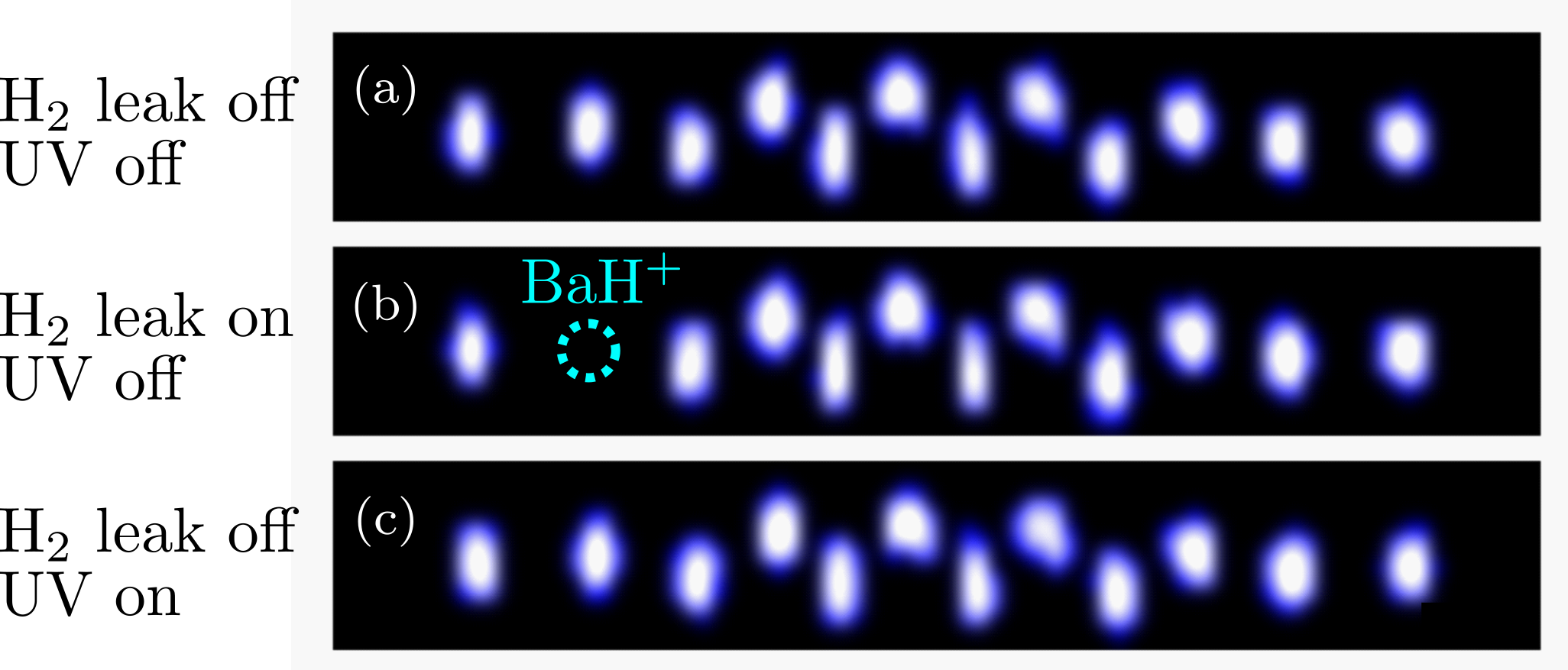}%
\vspace{-5pt}
\caption{\label{Image}
 The images of production and dissociation of BaH$^+$. (a) A 12-ion chain of $^{138}$Ba$^+$ is prepared with the UV photodissociation laser off. (b) H$_2$ is leaked in and a dark ion\textemdash~BaH$^+$ is produced in the chain labeled with a cyan dashed circle. (c) BaH$^+$ is photodissociated and returned to $^{138}$Ba$^+$.
\vspace{-4mm}}
\end{figure}

\begin{figure}[t]
\includegraphics[width=\linewidth]{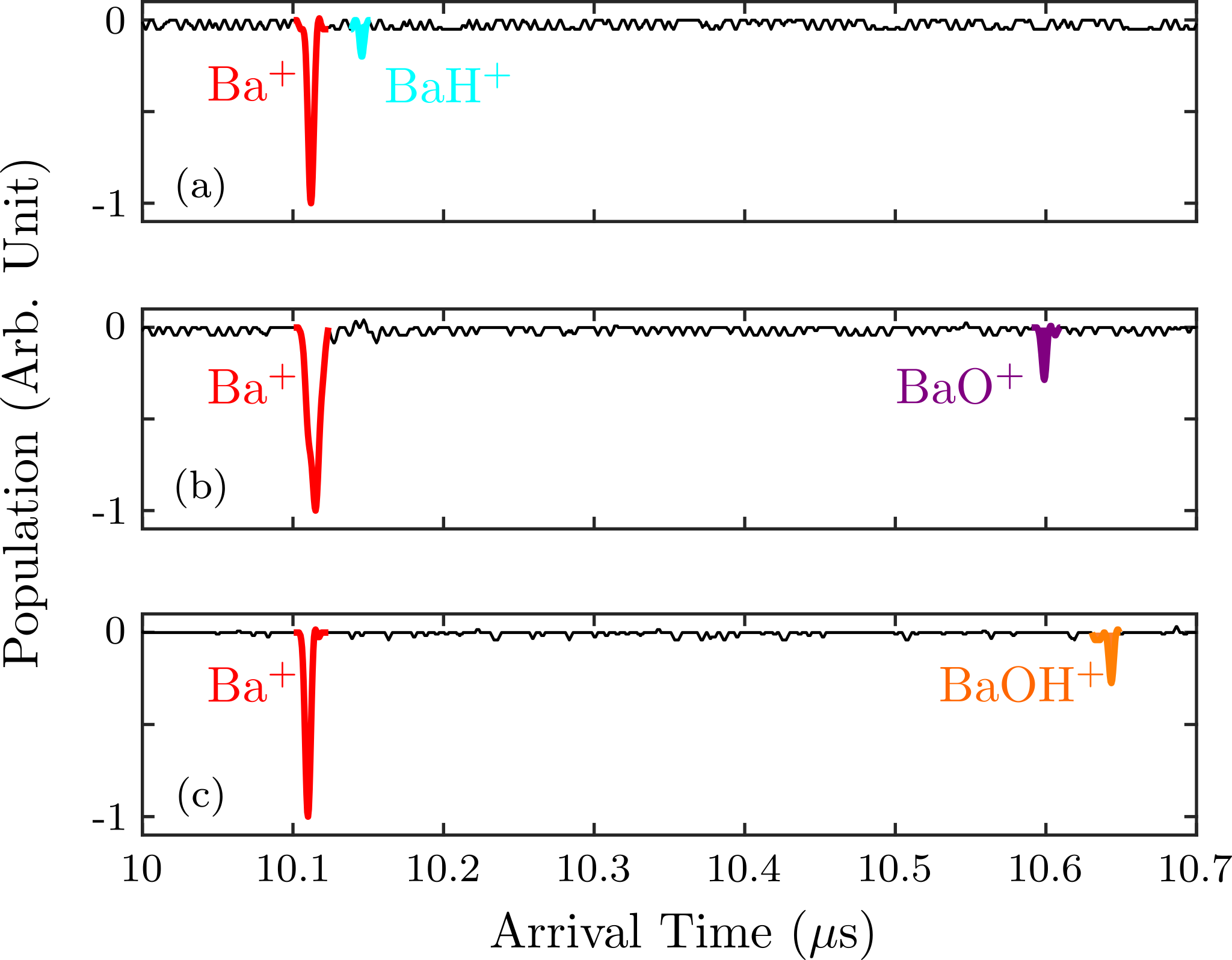}%
\vspace{-5pt}
\caption{\label{ToF}
Time of flight mass spectra (TOF-MS) of different molecular species. (a) BaH$^+$ starts to appear in a $^{138}$Ba$^+$ ion chain after H$_2$ is leaked in. (b) BaO$^+$ emerges in a $^{138}$Ba$^+$ ion chain after leaking in CO$_2$. (c) BaOH$^+$ is produced in a $^{138}$Ba$^+$ ion chain by introducing H$_2$O. 
\vspace{-4mm}}
\end{figure}

For the case of H$_2$, a small $^{138}$Ba$^+$ Coulomb crystal is prepared and  H$_2$ is introduced into the chamber with partial pressure of $\sim2\times$10$^{-8}$~mbar. 
Typically after approximately 30~mins, one of the ions in the crystal will become dark as shown in middle panel of Fig.~\ref{Image}.
At this point, the ions are ejected into the ToF-MS revealing that, as shown in Fig.\ref{ToF}(a), the dark ion is BaH$^+$, presumably produced via the reaction
\begin{equation}
\textrm{Ba}^+ + \textrm{H}_2 \rightarrow \textrm{BaH}^+ + \textrm{H}.
\label{BaH}
\end{equation}
As this reaction is endoergic by $\approx 2.1$~eV and $\approx 1.5$~eV for $^{138}$Ba$^+$ in $^2$S$_{1/2}$ and $^2$D$_{3/2}$ state, respectively, it likely proceeds via the $^{138}$Ba$^+$~($^2$P$_{1/2}$) electronic state accessed during laser cooling.
In this case, the reaction releases $\approx 0.4$~eV, of which 99.3\% is carried away by the light H atom, resulting in a BaH$^+$ molecular ion that remains embedded in the Coulomb crystal. 
This production pathway may be of use to experiments aiming to leverage the large rovibrational constants of BaH$^+$ to facilitate initialization into a single quantum state to search for the variation in m$_p$/m$_e$~\cite{Kajita_2011}.

The reaction with CO$_2$ is studied in a similar manner. 
CO$_2$ is introduced into the chamber at a partial pressure of $\sim 10^{-9}$~mbar. One ion in the crystal will react and turn dark typically within  5~mins. 
The reaction rate constant with CO$_2$ is around 100$\times$ higher than that of H$_2$. 
Ejection into the ToF-MS reveals that the dark ions are BaO$^+$, as shown in Fig.~\ref{ToF}(b), presumably formed from the reaction
\begin{equation}
\textrm{Ba}^+ + \textrm{CO}_2 \rightarrow \textrm{BaO}^+ + \textrm{CO}.
\label{BaO}
\end{equation}
This reaction can proceed for both the electronic $^2$P$_{1/2}$ and $^2$D$_{3/2}$ states of \barium, which release an energy of $\approx~1.1$~eV and $\approx~0.5$~eV, respectively~\cite{Murad1982,Roth2008}.  
Due to the heavier mass of the CO product, the BaO$^+$ can receive significantly more kinetic energy than the BaH$^+$.
The excess kinetic energy ($\leq 0.2$~eV) must be dissipated via sympathetic cooling from the remaining laser-cooled $^{138}$Ba$^+$ for the recoiling product ion to be recaptured in the crystal. 
Otherwise, the ion is likely to stay in a large orbit and experience heating due to the further collisions until it is eventually lost from the trap.

Finally, the reaction with H$_2$O is studied in a similar apparatus used for molecular ion spectroscopy~\cite{VMI}. 
After baking, the H$_2$O background pressure is suppressed below 10$^{-10}$~mbar.  
A small amount of H$_2$O can be released by ablating a barium chloride dihydrate (BaCl$_2$(H$_2$O)$_2$) pellet. 
Under this condition, one or more BaOH$^+$ are typically prepared in a $^{138}$Ba$^+$ ion chain, revealed by the ToF-MS.
These molecular ions are presumably formed from the reaction
\begin{equation}
\textrm{Ba}^+ + \textrm{H}_2\textrm{O} \rightarrow \textrm{BaOH}^+ + \textrm{H}.
\label{BaH}
\end{equation}
This reaction can proceed for the electronic $^2$S$_{1/2}$, $^2$P$_{1/2}$ and $^2$D$_{3/2}$ states of \barium, which release an energy of $\approx~0.4$~eV, $\approx~2.9$~eV, and $\approx~1.0$~eV, respectively, of which 99.4\% is carried away by the H product atom. 

Anecdotally, we observe that while the BaH$^+$/BaOH$^+$ product is maintained in the trap at low axial secular frequency ($2\pi\cdot9$~kHz), BaO$^+$ is only contained for higher axial secular frequency and/or within a 3D Coulomb crystal, which provides better sympathetic cooling than a linear ion chain. 
In fact at the maximum axial secular frequency here ($2\pi\cdot26$~kHz) we estimate the efficiency of recapturing the BaO$^+$ product in a 12-ion crystal is $\approx 50$\%. 
While the recoiling BaO$^+$ product does not have enough energy to escape the ion trap, it presumably resides in a large orbit that is relatively decoupled from the laser-cooled ion crystal. 
Here, further collisions or reactions can cause it to be lost from the trap. 
We occasionally observe the ion reappearing in the trap after several minutes. 
This suggests that with tighter trapping, and therefore improved sympathetic cooling from the remaining laser-cooled ions~\cite{Guggemos_2015}, all of the molecular ion reaction products can be maintained in the trap, thus opening the possibility of using photodissociation to reclaim the parent ion and avoid reloading the ion trap. 

To explore this possibility, it is necessary to determine potential photodissociation wavelengths for all three molecule systems. 
For BaH$^+$, Ref.~\cite{BaH+} 
suggests two useful paths for photodissociation.
First, a laser with wavelength $410(40)$~nm is able to drive BaH$^+$ from low-lying rovibrational states of the ground X$^1\Sigma^+$ state to the unbound A$^1\Sigma^+$ state.
Second, a laser with wavelength $245(20)$~nm can excite BaH$^+$ from the low-lying rovibrational states of the ground X$^1\Sigma^+$ state to the unbound C$^1\Sigma^+$ state~\cite{BaH+}.
Both pathways result in dissociation to a $^{138}$Ba$^+$ and H, potentially allowing the parent $^{138}$Ba$^+$ ion to be recovered.

For recovery of Ba$^+$ from BaO$^+$ via photodissociation, we perform \textit{ab initio} electronic structure calculations for the higher-energy excited states to predict transition energies to dissociative states.
The details of these calculations are given in a previous publication~\cite{Bartlett2015}, and we note here only the difference for the current study.  
In this case, we use the augmented valence triple zeta basis set for the O atom~\cite{Woon1993}.  
The lowest energy repulsive state is found to be 2$^2\Pi$, which can be reached from the ground state using light in the 210-228~nm range with a transition moment of approximately 0.47~D.

For recovery of Ba$^+$ from BaOH$^+$, the dissociation energy from a combined theoretical and experimental determination was reported to be within the range from 225 to 243~nm~\cite{Rossa2012}. 


Ideally, a single laser wavelength could be used to dissociate these products so that regardless of the background gas reactant the \barium can be recovered. 
Taking the intersection of the available photodissociation wavelengths suggests that light around 225~nm should be sufficient for recovering Ba$^+$ in all of the aforementioned cases.
Therefore, we introduce a single pulsed-dye laser beam at 225~nm (10~ns pulse, 3~mW/mm$^2$ on average) into the apparatus along the axial direction of the ion trap. 

A typical sequence for recovery of Ba$^+$ from, e.g., BaH$^+$ using this laser is shown in Fig.~\ref{Image}.
A 12-ion crystal is prepared and exposed to H$_2$ for $\approx$~30~mins before a BaH$^+$ ion appears, Fig.~\ref{Image}(b).
Next, the photodissociation laser is introduced and the BaH$^+$ is dissociated and the parent Ba$^+$ is recovered. 
We have also observed that a 370~nm continuous wave laser with a laser intensity of 2~mW/mm$^2$ recovers the Ba$^+$ from BaH$^+$. 
The pathway for dissociation with this laser is presumably a transition from the low-lying rovibrational states of X$^1\Sigma^+$ to A$^1\Sigma^+$ states above the dissociation threshold~\cite{BaH+}.

For BaO$^+$ and BaOH$^+$ we observe similar behavior.
Once dark ions appear, the photodissociation laser recovers the parent ion.
In the case of BaO$^+$, we observe that efficient photodissociation occurs for wavelengths ranging from 210~nm to 225~nm, suggesting that the technologically-convenient 5th harmonic of a Nd:YAG laser could be used in this case. 
For BaOH$^+$ we observe efficient photodissociation for wavelengths ranging from 225~nm to 240~nm. 

\begin{figure}[t]
\includegraphics[width=\linewidth]{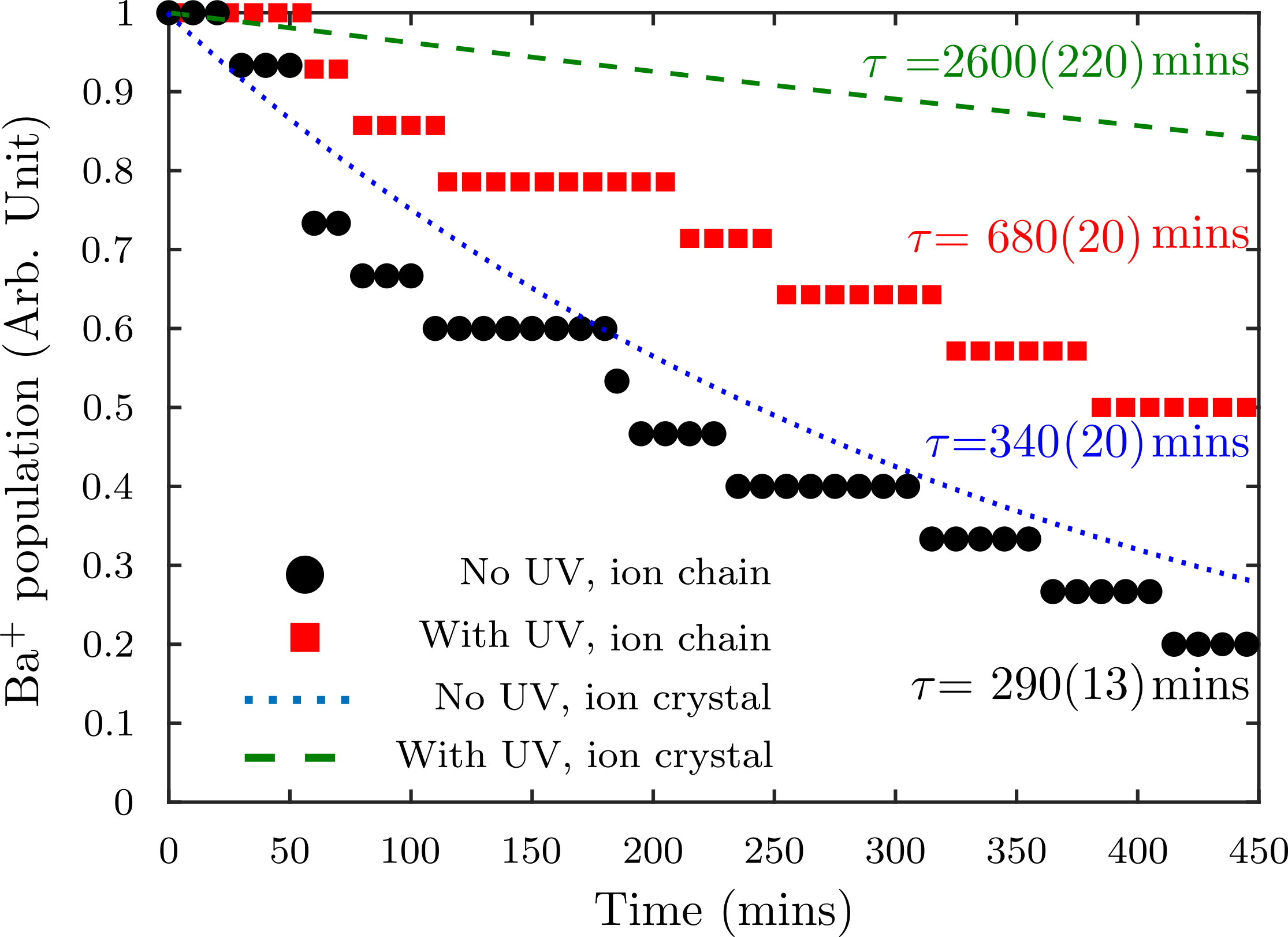}%
\vspace{-5pt}
\caption{\label{Lifetime}
The trapping lifetime of $^{138}$Ba$^+$ under different conditions. The red squares represent lifetime trajectory of a 12-ion chain of $^{138}$Ba$^+$  with photodissociation on; The black circles represent lifetime trajectory of a 12-ion chain of $^{138}$Ba$^+$ with photodissociation off; The green dash line represents lifetime trajectory of a 500-ion crystal of $^{138}$Ba$^+$ with photodissociation on;
The blue dot line represents lifetime trajectory of a 500-ion crystal of $^{138}$Ba$^+$ with photodissociation off. 
The relevant lifetimes are extracted from exponential fitting.
() denotes one standard error.
\vspace{-4mm}}
\end{figure}

With a single laser capable of dissociating the products of reactions with typical UHV gases, it is possible to extend the effective trap lifetime by periodically recovering Ba$^+$ from the produced molecular ions. 
To investigate this, the ion trap is illuminated with the photodissociation laser at 10~Hz repetition rate. 
Fig.~\ref{Lifetime} shows the lifetime enhancement of trapped $^{138}$Ba$^+$ due to the presence of the photodissociation laser. Once the photodissociation laser is off, the UHV background gas limits the lifetime of a 12-ion chain of $^{138}$Ba$^+$ to 290(13)~mins, represented by the black circles in Fig.~\ref{Lifetime}. 
The presence of the photodissociation laser extends the lifetime to 680(20)~mins, as indicated by the red squares in Fig.~\ref{Lifetime}. 

The lifetime enhancement is presumably limited by the $\approx 50$\% probability of capturing the BaO$^+$ products into the ion chain.  
To explore that, a $\sim$~500-ion crystal of $^{138}$Ba$^+$ is prepared, which can provide better sympathetic cooling and improve the trapping efficiency of recoiling product molecular ions. 
Without the UV dissociation laser present, the crystal represented by the dotted line in Fig.~\ref{Lifetime} shows a similar lifetime to that of the ion chain.
The dashed line in Fig.~\ref{Lifetime} shows the lifetime of the crystal is extended to 2600(220)~mins, further supporting the conclusion that the ion chain trap lifetime is limited by the efficiency of recapturing BaO$^+$.
As shown in Ref.~\cite{Guggemos_2015}, if the dark ions are produced in the trap center, the sympathetic cooling time is inversely proportional to the cube of the secular frequency. Thus, traps with higher secular frequency should expect significantly higher trap-lifetime improvements with the 225~nm photodissociation laser present.

Like Ba$^+$, other trapped ion species can also react with the common UHV background gases, e.g. H$_2$, CO, CO$_2$ and H$_2$O. 
Table~\ref{table1} shows a summary of possible reaction products and corresponding dissociation wavelengths for the products in their vibrational ground states. 

While only Ba$^+$ appears to have the fortuitous overlapping of all photodissociation pathways, the laser used for laser cooling of Ca$^+$ or Yb$^+$ dissociates one of the products``automatically'' meaning that only one additional laser is needed to dissociate all products.
Further, a properly baked UHV chamber will typically present a background gas primarily composed of H$_2$, CO, and CO$_2$, which would lead to only the hydride and oxide products.
As shown in column 4 of Tab.~\ref{table1}, there are technologically convenient pathways for dissociating these products for Mg$^+$, Ca$^+$, Ba$^+$, and Yb$^+$.

Of course, it is preferable to prevent these reactions from ever occurring.
While H$_2$ outgassing is typically assumed to be primarily responsible for ion loss as it tends to be the predominate residual gas, the reaction rate constants of CO and CO$_2$ may be considerably higher than that of H$_2$ -- e.g. here we find that the reaction with Ba$^+$ ions is around 100$\times$ faster for CO$_2$ than H$_2$ -- and all of these loss channels may play a significant role.
As a result, it is important to suppress the outgassing rates of H$_2$, CO, and CO$_2$.
In general, outgassing rates can be suppressed either by vacuum firing at $\approx 1300$~K or by an air bake at $\approx 700$~K, both of which lead to the building of an non-permeable oxide layer on the interior of the vacuum vessel~\cite{airbake}. 
Ref.~\cite{PhysRevSTAB.6.013201} demonstrates that the CO$_2$ desorption from the stainless steel chamber can also be suppressed through sputter coating of nonevaporable getter film followed by $\approx 600$~K baking.
\begin{table*}[h]
\begin{tabular}{c|c|c|c}
& Dissociation Wavelength & All products  & oxides \& hydrides\\
& (nm) & (nm) & (nm)\\
\cline{1-4} 
BeH$^+$ &140-210~\cite{BeH}/157~\cite{Brian2015}& & \\
BeO$^+$ &273-315~\cite{Ghalila_2008}&  N/A & N/A\\
BeOH$^+$ &223-236$^a$& & \\
\cline{1-4} 
MgH$^+$ &173-193/281~\cite{H_jbjerre_2009}&  &  \\
MgO$^+$ &156-176/368-478~\cite{Maatouk_2011}&  N/A & 173-176\\
MgOH$^+$ &324-383$^a$& & \\
\cline{1-4} 
CaH$^+$ &283-287~\cite{Kimura_2017}/370-421~\cite{Rugango2016}&   & Cooling laser + \\
CaO$^+$ &318-375~\cite{Khalil2013}& N/A & 318-375\\
CaOH$^+$ &233-255& & \\
\cline{1-4} 
SrH$^+$ & 240-270~\cite{ABUELKHER2021101264}& & \\
SrO$^+$ &208-226$^a$&  N/A & N/A\\
SrOH$^+$ &205-220$^a$& & \\
\cline{1-4} 
BaH$^+$ & 225-265/370-450~\cite{BaH+}& &  \\
BaO$^+$ &210-228~\cite{Bartlett2015}&  225-228 & 225-228\\
BaOH$^+$ &225-243~\cite{Rossa2012}&  & \\
\cline{1-4} 
YbH$^+$ &369~\cite{Sugiyama1997}/405~\cite{Hoang2020}&Cooling laser + &Cooling laser +  \\
YbO$^+$ &188-200/257-286$^a$ &257-286$^a$ &257-286$^a$ \\
YbOH$^+$ &216-330$^a$&&\\

\hline
\end{tabular}
    \begin{tablenotes}
      \small
      \item a. the relevant dissociation wavelengths are estimated in this work.
    \end{tablenotes}
    \caption{A list of possible molecular ions produced by  chemical reaction with UHV residual gas (CO$_2$, CO, H$_2$O and H$_2$) and corresponding dissociation wavelengths. Intersection means the common wavelength to dissociate all the relevant complexes. 
   }
    \label{table1}   

\end{table*}


In summary, we observed that the trap lifetime of Ba$^+$ ions is limited by chemical reaction with some common UHV background gases (H$_2$, CO, CO$_2$ and H$_2$O) once the micromotion-interruption    -induced loss is suppressed with high-enough trap depth. 
Using relevant spectroscopic data and \textit{ab initio} molecular structure calculations, we identified photodissociation pathways for all of the produced molecular ions and recovered the parent atomic ion from the product molecular ions with a single laser.
Specifically, we observed an improvement in the effective trap lifetime by $2.3(1)\times$ and $7.6(8)\times$ for an ion chain and ion crystal, respectively, with a 225~nm laser.
The data suggest that the lifetime enhancement is limited by the efficiency of BaO$^+$ recapture, which should improve significantly with a larger secular frequency as employed in most trapped ion quantum logic experiments.
This process should be straightforward to extend to other atomic ion species and the requisite dissociation wavelengths are provided.
This technique could be employed in trapped ion quantum logic experiments to recover lost ions and avoid reloading of the ion trap.

\section*{Acknowledgements}
We acknowledge support from the NSF QLCI program through grant number OMA-2016245.
This work was also supported in part by National Science
Foundation (Grants No. PHY-1255526, No. PHY-1415560, No. PHY-1912555, No. CHE-1900555,
and No. DGE-1650604) and Army Research Office (Grants
No. W911NF-15-1-0121, No. W911NF-14-1-0378,
No. W911NF-13-1-0213 and  W911NF-17-1-0071) grants.

\bibliographystyle{apsrev4-1_no_Arxiv}
\bibliography{allrefs}

\begin{thebibliography}{41}%
\makeatletter
\providecommand \@ifxundefined [1]{%
 \@ifx{#1\undefined}
}%
\providecommand \@ifnum [1]{%
 \ifnum #1\expandafter \@firstoftwo
 \else \expandafter \@secondoftwo
 \fi
}%
\providecommand \@ifx [1]{%
 \ifx #1\expandafter \@firstoftwo
 \else \expandafter \@secondoftwo
 \fi
}%
\providecommand \natexlab [1]{#1}%
\providecommand \enquote  [1]{``#1''}%
\providecommand \bibnamefont  [1]{#1}%
\providecommand \bibfnamefont [1]{#1}%
\providecommand \citenamefont [1]{#1}%
\providecommand \href@noop [0]{\@secondoftwo}%
\providecommand \href [0]{\begingroup \@sanitize@url \@href}%
\providecommand \@href[1]{\@@startlink{#1}\@@href}%
\providecommand \@@href[1]{\endgroup#1\@@endlink}%
\providecommand \@sanitize@url [0]{\catcode `\\12\catcode `\$12\catcode
  `\&12\catcode `\#12\catcode `\^12\catcode `\_12\catcode `\%12\relax}%
\providecommand \@@startlink[1]{}%
\providecommand \@@endlink[0]{}%
\providecommand \url  [0]{\begingroup\@sanitize@url \@url }%
\providecommand \@url [1]{\endgroup\@href {#1}{\urlprefix }}%
\providecommand \urlprefix  [0]{URL }%
\providecommand \Eprint [0]{\href }%
\providecommand \doibase [0]{http://dx.doi.org/}%
\providecommand \selectlanguage [0]{\@gobble}%
\providecommand \bibinfo  [0]{\@secondoftwo}%
\providecommand \bibfield  [0]{\@secondoftwo}%
\providecommand \translation [1]{[#1]}%
\providecommand \BibitemOpen [0]{}%
\providecommand \bibitemStop [0]{}%
\providecommand \bibitemNoStop [0]{.\EOS\space}%
\providecommand \EOS [0]{\spacefactor3000\relax}%
\providecommand \BibitemShut  [1]{\csname bibitem#1\endcsname}%
\let\auto@bib@innerbib\@empty
\bibitem [{\citenamefont {Bruzewicz}\ \emph
  {et~al.}(2019{\natexlab{a}})\citenamefont {Bruzewicz}, \citenamefont
  {Chiaverini}, \citenamefont {McConnell},\ and\ \citenamefont
  {Sage}}]{doi:10.1063/1.5088164}%
  \BibitemOpen
  \bibfield  {author} {\bibinfo {author} {\bibfnamefont {C.~D.}\ \bibnamefont
  {Bruzewicz}}, \bibinfo {author} {\bibfnamefont {J.}~\bibnamefont
  {Chiaverini}}, \bibinfo {author} {\bibfnamefont {R.}~\bibnamefont
  {McConnell}}, \ and\ \bibinfo {author} {\bibfnamefont {J.~M.}\ \bibnamefont
  {Sage}},\ }\href {\doibase 10.1063/1.5088164} {\bibfield  {journal} {\bibinfo
   {journal} {Applied Physics Reviews}\ }\textbf {\bibinfo {volume} {6}},\
  \bibinfo {pages} {021314} (\bibinfo {year} {2019}{\natexlab{a}})}\BibitemShut
  {NoStop}%
\bibitem [{\citenamefont {Christensen}\ \emph {et~al.}(2020)\citenamefont
  {Christensen}, \citenamefont {Hucul}, \citenamefont {Campbell},\ and\
  \citenamefont {Hudson}}]{Christensen2020}%
  \BibitemOpen
  \bibfield  {author} {\bibinfo {author} {\bibfnamefont {J.~E.}\ \bibnamefont
  {Christensen}}, \bibinfo {author} {\bibfnamefont {D.}~\bibnamefont {Hucul}},
  \bibinfo {author} {\bibfnamefont {W.~C.}\ \bibnamefont {Campbell}}, \ and\
  \bibinfo {author} {\bibfnamefont {E.~R.}\ \bibnamefont {Hudson}},\ }\href
  {\doibase 10.1038/s41534-020-0265-5} {\bibfield  {journal} {\bibinfo
  {journal} {npj Quantum Information}\ }\textbf {\bibinfo {volume} {6}},\
  \bibinfo {pages} {35} (\bibinfo {year} {2020})}\BibitemShut {NoStop}%
\bibitem [{\citenamefont {Ballance}\ \emph {et~al.}(2016)\citenamefont
  {Ballance}, \citenamefont {Harty}, \citenamefont {Linke}, \citenamefont
  {Sepiol},\ and\ \citenamefont {Lucas}}]{Ballance2016}%
  \BibitemOpen
  \bibfield  {author} {\bibinfo {author} {\bibfnamefont {C.~J.}\ \bibnamefont
  {Ballance}}, \bibinfo {author} {\bibfnamefont {T.~P.}\ \bibnamefont {Harty}},
  \bibinfo {author} {\bibfnamefont {N.~M.}\ \bibnamefont {Linke}}, \bibinfo
  {author} {\bibfnamefont {M.~A.}\ \bibnamefont {Sepiol}}, \ and\ \bibinfo
  {author} {\bibfnamefont {D.~M.}\ \bibnamefont {Lucas}},\ }\href {\doibase
  10.1103/PhysRevLett.117.060504} {\bibfield  {journal} {\bibinfo  {journal}
  {Phys. Rev. Lett.}\ }\textbf {\bibinfo {volume} {117}},\ \bibinfo {pages}
  {060504} (\bibinfo {year} {2016})}\BibitemShut {NoStop}%
\bibitem [{\citenamefont {Srinivas}\ \emph {et~al.}()\citenamefont {Srinivas}
  \emph {et~al.}}]{Srinivas2021}%
  \BibitemOpen
  \bibfield  {author} {\bibinfo {author} {\bibfnamefont {R.}~\bibnamefont
  {Srinivas}} \emph {et~al.},\ }\href@noop {} {\bibinfo  {journal}
  {arXiv:2102.12533}\ }\BibitemShut {NoStop}%
\bibitem [{\citenamefont {Gaebler}\ \emph {et~al.}(2016)\citenamefont
  {Gaebler}, \citenamefont {Tan}, \citenamefont {Lin}, \citenamefont {Wan},
  \citenamefont {Bowler}, \citenamefont {Keith}, \citenamefont {Glancy},
  \citenamefont {Coakley}, \citenamefont {Knill}, \citenamefont {Leibfried},\
  and\ \citenamefont {Wineland}}]{Gaebler2016}%
  \BibitemOpen
\bibfield  {journal} {  }\bibfield  {author} {\bibinfo {author} {\bibfnamefont
  {J.~P.}\ \bibnamefont {Gaebler}}, \bibinfo {author} {\bibfnamefont {T.~R.}\
  \bibnamefont {Tan}}, \bibinfo {author} {\bibfnamefont {Y.}~\bibnamefont
  {Lin}}, \bibinfo {author} {\bibfnamefont {Y.}~\bibnamefont {Wan}}, \bibinfo
  {author} {\bibfnamefont {R.}~\bibnamefont {Bowler}}, \bibinfo {author}
  {\bibfnamefont {A.~C.}\ \bibnamefont {Keith}}, \bibinfo {author}
  {\bibfnamefont {S.}~\bibnamefont {Glancy}}, \bibinfo {author} {\bibfnamefont
  {K.}~\bibnamefont {Coakley}}, \bibinfo {author} {\bibfnamefont
  {E.}~\bibnamefont {Knill}}, \bibinfo {author} {\bibfnamefont
  {D.}~\bibnamefont {Leibfried}}, \ and\ \bibinfo {author} {\bibfnamefont
  {D.~J.}\ \bibnamefont {Wineland}},\ }\href {\doibase
  10.1103/PhysRevLett.117.060505} {\bibfield  {journal} {\bibinfo  {journal}
  {Phys. Rev. Lett.}\ }\textbf {\bibinfo {volume} {117}},\ \bibinfo {pages}
  {060505} (\bibinfo {year} {2016})}\BibitemShut {NoStop}%
\bibitem [{\citenamefont {Bruzewicz}\ \emph
  {et~al.}(2019{\natexlab{b}})\citenamefont {Bruzewicz}, \citenamefont
  {McConnell}, \citenamefont {Stuart}, \citenamefont {Sage},\ and\
  \citenamefont {Chiaverini}}]{Bruzewicz2019}%
  \BibitemOpen
  \bibfield  {author} {\bibinfo {author} {\bibfnamefont {C.~D.}\ \bibnamefont
  {Bruzewicz}}, \bibinfo {author} {\bibfnamefont {R.}~\bibnamefont
  {McConnell}}, \bibinfo {author} {\bibfnamefont {J.}~\bibnamefont {Stuart}},
  \bibinfo {author} {\bibfnamefont {J.~M.}\ \bibnamefont {Sage}}, \ and\
  \bibinfo {author} {\bibfnamefont {J.}~\bibnamefont {Chiaverini}},\ }\href
  {\doibase 10.1038/s41534-019-0218-z} {\bibfield  {journal} {\bibinfo
  {journal} {npj Quantum Information}\ }\textbf {\bibinfo {volume} {5}},\
  \bibinfo {pages} {102} (\bibinfo {year} {2019}{\natexlab{b}})}\BibitemShut
  {NoStop}%
\bibitem [{\citenamefont {Mehta}\ \emph {et~al.}(2020)\citenamefont {Mehta},
  \citenamefont {Zhang}, \citenamefont {Malinowski}, \citenamefont {Nguyen},
  \citenamefont {Stadler},\ and\ \citenamefont {Home}}]{Mehta2020}%
  \BibitemOpen
  \bibfield  {author} {\bibinfo {author} {\bibfnamefont {K.~K.}\ \bibnamefont
  {Mehta}}, \bibinfo {author} {\bibfnamefont {C.}~\bibnamefont {Zhang}},
  \bibinfo {author} {\bibfnamefont {M.}~\bibnamefont {Malinowski}}, \bibinfo
  {author} {\bibfnamefont {T.-L.}\ \bibnamefont {Nguyen}}, \bibinfo {author}
  {\bibfnamefont {M.}~\bibnamefont {Stadler}}, \ and\ \bibinfo {author}
  {\bibfnamefont {J.~P.}\ \bibnamefont {Home}},\ }\href {\doibase
  10.1038/s41586-020-2823-6} {\bibfield  {journal} {\bibinfo  {journal}
  {Nature}\ }\textbf {\bibinfo {volume} {586}},\ \bibinfo {pages} {533}
  (\bibinfo {year} {2020})}\BibitemShut {NoStop}%
\bibitem [{\citenamefont {Bruzewicz}\ \emph {et~al.}(2016)\citenamefont
  {Bruzewicz}, \citenamefont {McConnell}, \citenamefont {Chiaverini},\ and\
  \citenamefont {Sage}}]{Bruzewicz2016}%
  \BibitemOpen
  \bibfield  {author} {\bibinfo {author} {\bibfnamefont {C.~D.}\ \bibnamefont
  {Bruzewicz}}, \bibinfo {author} {\bibfnamefont {R.}~\bibnamefont
  {McConnell}}, \bibinfo {author} {\bibfnamefont {J.}~\bibnamefont
  {Chiaverini}}, \ and\ \bibinfo {author} {\bibfnamefont {J.~M.}\ \bibnamefont
  {Sage}},\ }\href {\doibase 10.1038/ncomms13005} {\bibfield  {journal}
  {\bibinfo  {journal} {Nature Communications}\ }\textbf {\bibinfo {volume}
  {7}},\ \bibinfo {pages} {13005} (\bibinfo {year} {2016})}\BibitemShut
  {NoStop}%
\bibitem [{\citenamefont {Pagano}\ \emph {et~al.}(2018)\citenamefont {Pagano},
  \citenamefont {Hess}, \citenamefont {Kaplan}, \citenamefont {Tan},
  \citenamefont {Richerme}, \citenamefont {Becker}, \citenamefont
  {Kyprianidis}, \citenamefont {Zhang}, \citenamefont {Birckelbaw},
  \citenamefont {Hernandez}, \citenamefont {Wu},\ and\ \citenamefont
  {Monroe}}]{Pagano_2018}%
  \BibitemOpen
  \bibfield  {author} {\bibinfo {author} {\bibfnamefont {G.}~\bibnamefont
  {Pagano}}, \bibinfo {author} {\bibfnamefont {P.~W.}\ \bibnamefont {Hess}},
  \bibinfo {author} {\bibfnamefont {H.~B.}\ \bibnamefont {Kaplan}}, \bibinfo
  {author} {\bibfnamefont {W.~L.}\ \bibnamefont {Tan}}, \bibinfo {author}
  {\bibfnamefont {P.}~\bibnamefont {Richerme}}, \bibinfo {author}
  {\bibfnamefont {P.}~\bibnamefont {Becker}}, \bibinfo {author} {\bibfnamefont
  {A.}~\bibnamefont {Kyprianidis}}, \bibinfo {author} {\bibfnamefont
  {J.}~\bibnamefont {Zhang}}, \bibinfo {author} {\bibfnamefont
  {E.}~\bibnamefont {Birckelbaw}}, \bibinfo {author} {\bibfnamefont {M.~R.}\
  \bibnamefont {Hernandez}}, \bibinfo {author} {\bibfnamefont {Y.}~\bibnamefont
  {Wu}}, \ and\ \bibinfo {author} {\bibfnamefont {C.}~\bibnamefont {Monroe}},\
  }\href {\doibase 10.1088/2058-9565/aae0fe} {\bibfield  {journal} {\bibinfo
  {journal} {Quantum Science and Technology}\ }\textbf {\bibinfo {volume}
  {4}},\ \bibinfo {pages} {014004} (\bibinfo {year} {2018})}\BibitemShut
  {NoStop}%
\bibitem [{\citenamefont {Brandl}\ \emph {et~al.}(2016)\citenamefont {Brandl},
  \citenamefont {van Mourik}, \citenamefont {Postler}, \citenamefont {Nolf},
  \citenamefont {Lakhmanskiy}, \citenamefont {Paiva}, \citenamefont {Möller},
  \citenamefont {Daniilidis}, \citenamefont {Häffner}, \citenamefont
  {Kaushal}, \citenamefont {Ruster}, \citenamefont {Warschburger},
  \citenamefont {Kaufmann}, \citenamefont {Poschinger}, \citenamefont
  {Schmidt-Kaler}, \citenamefont {Schindler}, \citenamefont {Monz},\ and\
  \citenamefont {Blatt}}]{Brandl2016}%
  \BibitemOpen
  \bibfield  {author} {\bibinfo {author} {\bibfnamefont {M.~F.}\ \bibnamefont
  {Brandl}}, \bibinfo {author} {\bibfnamefont {M.~W.}\ \bibnamefont {van
  Mourik}}, \bibinfo {author} {\bibfnamefont {L.}~\bibnamefont {Postler}},
  \bibinfo {author} {\bibfnamefont {A.}~\bibnamefont {Nolf}}, \bibinfo {author}
  {\bibfnamefont {K.}~\bibnamefont {Lakhmanskiy}}, \bibinfo {author}
  {\bibfnamefont {R.~R.}\ \bibnamefont {Paiva}}, \bibinfo {author}
  {\bibfnamefont {S.}~\bibnamefont {Möller}}, \bibinfo {author} {\bibfnamefont
  {N.}~\bibnamefont {Daniilidis}}, \bibinfo {author} {\bibfnamefont
  {H.}~\bibnamefont {Häffner}}, \bibinfo {author} {\bibfnamefont
  {V.}~\bibnamefont {Kaushal}}, \bibinfo {author} {\bibfnamefont
  {T.}~\bibnamefont {Ruster}}, \bibinfo {author} {\bibfnamefont
  {C.}~\bibnamefont {Warschburger}}, \bibinfo {author} {\bibfnamefont
  {H.}~\bibnamefont {Kaufmann}}, \bibinfo {author} {\bibfnamefont {U.~G.}\
  \bibnamefont {Poschinger}}, \bibinfo {author} {\bibfnamefont
  {F.}~\bibnamefont {Schmidt-Kaler}}, \bibinfo {author} {\bibfnamefont
  {P.}~\bibnamefont {Schindler}}, \bibinfo {author} {\bibfnamefont
  {T.}~\bibnamefont {Monz}}, \ and\ \bibinfo {author} {\bibfnamefont
  {R.}~\bibnamefont {Blatt}},\ }\href {\doibase 10.1063/1.4966970} {\bibfield
  {journal} {\bibinfo  {journal} {Review of Scientific Instruments}\ }\textbf
  {\bibinfo {volume} {87}},\ \bibinfo {pages} {113103} (\bibinfo {year}
  {2016})}\BibitemShut {NoStop}%
\bibitem [{\citenamefont {Leopold}\ \emph {et~al.}(2019)\citenamefont
  {Leopold}, \citenamefont {King}, \citenamefont {Micke}, \citenamefont
  {Bautista-Salvador}, \citenamefont {Heip}, \citenamefont {Ospelkaus},
  \citenamefont {Crespo López-Urrutia},\ and\ \citenamefont
  {Schmidt}}]{Leopold2019}%
  \BibitemOpen
  \bibfield  {author} {\bibinfo {author} {\bibfnamefont {T.}~\bibnamefont
  {Leopold}}, \bibinfo {author} {\bibfnamefont {S.~A.}\ \bibnamefont {King}},
  \bibinfo {author} {\bibfnamefont {P.}~\bibnamefont {Micke}}, \bibinfo
  {author} {\bibfnamefont {A.}~\bibnamefont {Bautista-Salvador}}, \bibinfo
  {author} {\bibfnamefont {J.~C.}\ \bibnamefont {Heip}}, \bibinfo {author}
  {\bibfnamefont {C.}~\bibnamefont {Ospelkaus}}, \bibinfo {author}
  {\bibfnamefont {J.~R.}\ \bibnamefont {Crespo López-Urrutia}}, \ and\
  \bibinfo {author} {\bibfnamefont {P.~O.}\ \bibnamefont {Schmidt}},\ }\href
  {\doibase 10.1063/1.5100594} {\bibfield  {journal} {\bibinfo  {journal}
  {Review of Scientific Instruments}\ }\textbf {\bibinfo {volume} {90}},\
  \bibinfo {pages} {073201} (\bibinfo {year} {2019})}\BibitemShut {NoStop}%
\bibitem [{\citenamefont {Sawyer}\ \emph {et~al.}(2015)\citenamefont {Sawyer},
  \citenamefont {Bohnet}, \citenamefont {Britton},\ and\ \citenamefont
  {Bollinger}}]{Brian2015}%
  \BibitemOpen
  \bibfield  {author} {\bibinfo {author} {\bibfnamefont {B.~C.}\ \bibnamefont
  {Sawyer}}, \bibinfo {author} {\bibfnamefont {J.~G.}\ \bibnamefont {Bohnet}},
  \bibinfo {author} {\bibfnamefont {J.~W.}\ \bibnamefont {Britton}}, \ and\
  \bibinfo {author} {\bibfnamefont {J.~J.}\ \bibnamefont {Bollinger}},\ }\href
  {\doibase 10.1103/PhysRevA.91.011401} {\bibfield  {journal} {\bibinfo
  {journal} {Phys. Rev. A}\ }\textbf {\bibinfo {volume} {91}},\ \bibinfo
  {pages} {011401} (\bibinfo {year} {2015})}\BibitemShut {NoStop}%
\bibitem [{\citenamefont {Chen}\ \emph {et~al.}(2011)\citenamefont {Chen},
  \citenamefont {Schowalter}, \citenamefont {Kotochigova}, \citenamefont
  {Petrov}, \citenamefont {Rellergert}, \citenamefont {Sullivan},\ and\
  \citenamefont {Hudson}}]{Kuang2011}%
  \BibitemOpen
  \bibfield  {author} {\bibinfo {author} {\bibfnamefont {K.}~\bibnamefont
  {Chen}}, \bibinfo {author} {\bibfnamefont {S.~J.}\ \bibnamefont
  {Schowalter}}, \bibinfo {author} {\bibfnamefont {S.}~\bibnamefont
  {Kotochigova}}, \bibinfo {author} {\bibfnamefont {A.}~\bibnamefont {Petrov}},
  \bibinfo {author} {\bibfnamefont {W.~G.}\ \bibnamefont {Rellergert}},
  \bibinfo {author} {\bibfnamefont {S.~T.}\ \bibnamefont {Sullivan}}, \ and\
  \bibinfo {author} {\bibfnamefont {E.~R.}\ \bibnamefont {Hudson}},\ }\href
  {\doibase 10.1103/PhysRevA.83.030501} {\bibfield  {journal} {\bibinfo
  {journal} {Phys. Rev. A}\ }\textbf {\bibinfo {volume} {83}},\ \bibinfo
  {pages} {030501} (\bibinfo {year} {2011})}\BibitemShut {NoStop}%
\bibitem [{\citenamefont {Guggemos}\ \emph {et~al.}(2015)\citenamefont
  {Guggemos}, \citenamefont {Heinrich}, \citenamefont {Herrera-Sancho},
  \citenamefont {Blatt},\ and\ \citenamefont {Roos}}]{Guggemos_2015}%
  \BibitemOpen
  \bibfield  {author} {\bibinfo {author} {\bibfnamefont {M.}~\bibnamefont
  {Guggemos}}, \bibinfo {author} {\bibfnamefont {D.}~\bibnamefont {Heinrich}},
  \bibinfo {author} {\bibfnamefont {O.~A.}\ \bibnamefont {Herrera-Sancho}},
  \bibinfo {author} {\bibfnamefont {R.}~\bibnamefont {Blatt}}, \ and\ \bibinfo
  {author} {\bibfnamefont {C.~F.}\ \bibnamefont {Roos}},\ }\href {\doibase
  10.1088/1367-2630/17/10/103001} {\bibfield  {journal} {\bibinfo  {journal}
  {New Journal of Physics}\ }\textbf {\bibinfo {volume} {17}},\ \bibinfo
  {pages} {103001} (\bibinfo {year} {2015})}\BibitemShut {NoStop}%
\bibitem [{\citenamefont {Chen}\ \emph {et~al.}(2014)\citenamefont {Chen},
  \citenamefont {Sullivan},\ and\ \citenamefont {Hudson}}]{Kuang143009}%
  \BibitemOpen
  \bibfield  {author} {\bibinfo {author} {\bibfnamefont {K.}~\bibnamefont
  {Chen}}, \bibinfo {author} {\bibfnamefont {S.~T.}\ \bibnamefont {Sullivan}},
  \ and\ \bibinfo {author} {\bibfnamefont {E.~R.}\ \bibnamefont {Hudson}},\
  }\href {\doibase 10.1103/PhysRevLett.112.143009} {\bibfield  {journal}
  {\bibinfo  {journal} {Phys. Rev. Lett.}\ }\textbf {\bibinfo {volume} {112}},\
  \bibinfo {pages} {143009} (\bibinfo {year} {2014})}\BibitemShut {NoStop}%
\bibitem [{\citenamefont {Chen}\ \emph {et~al.}(2013)\citenamefont {Chen},
  \citenamefont {Sullivan}, \citenamefont {Rellergert},\ and\ \citenamefont
  {Hudson}}]{Kuang173003}%
  \BibitemOpen
  \bibfield  {author} {\bibinfo {author} {\bibfnamefont {K.}~\bibnamefont
  {Chen}}, \bibinfo {author} {\bibfnamefont {S.~T.}\ \bibnamefont {Sullivan}},
  \bibinfo {author} {\bibfnamefont {W.~G.}\ \bibnamefont {Rellergert}}, \ and\
  \bibinfo {author} {\bibfnamefont {E.~R.}\ \bibnamefont {Hudson}},\ }\href
  {\doibase 10.1103/PhysRevLett.110.173003} {\bibfield  {journal} {\bibinfo
  {journal} {Phys. Rev. Lett.}\ }\textbf {\bibinfo {volume} {110}},\ \bibinfo
  {pages} {173003} (\bibinfo {year} {2013})}\BibitemShut {NoStop}%
\bibitem [{\citenamefont {Schowalter}\ \emph {et~al.}(2016)\citenamefont
  {Schowalter}, \citenamefont {Dunning}, \citenamefont {Chen}, \citenamefont
  {Puri}, \citenamefont {Schneider},\ and\ \citenamefont
  {Hudson}}]{Bifurcation}%
  \BibitemOpen
  \bibfield  {author} {\bibinfo {author} {\bibfnamefont {S.~J.}\ \bibnamefont
  {Schowalter}}, \bibinfo {author} {\bibfnamefont {A.~J.}\ \bibnamefont
  {Dunning}}, \bibinfo {author} {\bibfnamefont {K.}~\bibnamefont {Chen}},
  \bibinfo {author} {\bibfnamefont {P.}~\bibnamefont {Puri}}, \bibinfo {author}
  {\bibfnamefont {C.}~\bibnamefont {Schneider}}, \ and\ \bibinfo {author}
  {\bibfnamefont {E.~R.}\ \bibnamefont {Hudson}},\ }\href {\doibase
  10.1038/ncomms12448} {\bibfield  {journal} {\bibinfo  {journal} {Nature
  Communications}\ }\textbf {\bibinfo {volume} {7}},\ \bibinfo {pages} {12448}
  (\bibinfo {year} {2016})}\BibitemShut {NoStop}%
\bibitem [{\citenamefont {Cetina}\ \emph {et~al.}(2012)\citenamefont {Cetina},
  \citenamefont {Grier},\ and\ \citenamefont {Vuleti\ifmmode~\acute{c}\else
  \'{c}\fi{}}}]{Cetina2012}%
  \BibitemOpen
  \bibfield  {author} {\bibinfo {author} {\bibfnamefont {M.}~\bibnamefont
  {Cetina}}, \bibinfo {author} {\bibfnamefont {A.~T.}\ \bibnamefont {Grier}}, \
  and\ \bibinfo {author} {\bibfnamefont {V.}~\bibnamefont
  {Vuleti\ifmmode~\acute{c}\else \'{c}\fi{}}},\ }\href {\doibase
  10.1103/PhysRevLett.109.253201} {\bibfield  {journal} {\bibinfo  {journal}
  {Phys. Rev. Lett.}\ }\textbf {\bibinfo {volume} {109}},\ \bibinfo {pages}
  {253201} (\bibinfo {year} {2012})}\BibitemShut {NoStop}%
\bibitem [{\citenamefont {Mills}(2020)}]{MikeThesis}%
  \BibitemOpen
  \bibfield  {author} {\bibinfo {author} {\bibfnamefont {M.}~\bibnamefont
  {Mills}},\ }\href@noop {} {Ph.D. thesis},\ \bibinfo  {school} {University of
  California Los Angeles} (\bibinfo {year} {2020})\BibitemShut {NoStop}%
\bibitem [{\citenamefont {Home}\ \emph {et~al.}(2011)\citenamefont {Home},
  \citenamefont {Hanneke}, \citenamefont {Jost}, \citenamefont {Leibfried},\
  and\ \citenamefont {Wineland}}]{Home_2011}%
  \BibitemOpen
  \bibfield  {author} {\bibinfo {author} {\bibfnamefont {J.~P.}\ \bibnamefont
  {Home}}, \bibinfo {author} {\bibfnamefont {D.}~\bibnamefont {Hanneke}},
  \bibinfo {author} {\bibfnamefont {J.~D.}\ \bibnamefont {Jost}}, \bibinfo
  {author} {\bibfnamefont {D.}~\bibnamefont {Leibfried}}, \ and\ \bibinfo
  {author} {\bibfnamefont {D.~J.}\ \bibnamefont {Wineland}},\ }\href {\doibase
  10.1088/1367-2630/13/7/073026} {\bibfield  {journal} {\bibinfo  {journal}
  {New Journal of Physics}\ }\textbf {\bibinfo {volume} {13}},\ \bibinfo
  {pages} {073026} (\bibinfo {year} {2011})}\BibitemShut {NoStop}%
\bibitem [{\citenamefont {Roth}\ \emph {et~al.}(2008)\citenamefont {Roth},
  \citenamefont {Offenberg}, \citenamefont {Zhang},\ and\ \citenamefont
  {Schiller}}]{Roth2008}%
  \BibitemOpen
  \bibfield  {author} {\bibinfo {author} {\bibfnamefont {B.}~\bibnamefont
  {Roth}}, \bibinfo {author} {\bibfnamefont {D.}~\bibnamefont {Offenberg}},
  \bibinfo {author} {\bibfnamefont {C.~B.}\ \bibnamefont {Zhang}}, \ and\
  \bibinfo {author} {\bibfnamefont {S.}~\bibnamefont {Schiller}},\ }\href
  {\doibase 10.1103/PhysRevA.78.042709} {\bibfield  {journal} {\bibinfo
  {journal} {Phys. Rev. A}\ }\textbf {\bibinfo {volume} {78}},\ \bibinfo
  {pages} {042709} (\bibinfo {year} {2008})}\BibitemShut {NoStop}%
\bibitem [{\citenamefont {Schowalter}\ \emph {et~al.}(2012)\citenamefont
  {Schowalter}, \citenamefont {Chen}, \citenamefont {Rellergert}, \citenamefont
  {Sullivan},\ and\ \citenamefont {Hudson}}]{TOF}%
  \BibitemOpen
  \bibfield  {author} {\bibinfo {author} {\bibfnamefont {S.~J.}\ \bibnamefont
  {Schowalter}}, \bibinfo {author} {\bibfnamefont {K.}~\bibnamefont {Chen}},
  \bibinfo {author} {\bibfnamefont {W.~G.}\ \bibnamefont {Rellergert}},
  \bibinfo {author} {\bibfnamefont {S.~T.}\ \bibnamefont {Sullivan}}, \ and\
  \bibinfo {author} {\bibfnamefont {E.~R.}\ \bibnamefont {Hudson}},\ }\href
  {\doibase 10.1063/1.3700216} {\bibfield  {journal} {\bibinfo  {journal}
  {Review of Scientific Instruments}\ }\textbf {\bibinfo {volume} {83}},\
  \bibinfo {pages} {043103} (\bibinfo {year} {2012})}\BibitemShut {NoStop}%
\bibitem [{\citenamefont {Kajita}\ \emph {et~al.}(2011)\citenamefont {Kajita},
  \citenamefont {Abe}, \citenamefont {Hada},\ and\ \citenamefont
  {Moriwaki}}]{Kajita_2011}%
  \BibitemOpen
  \bibfield  {author} {\bibinfo {author} {\bibfnamefont {M.}~\bibnamefont
  {Kajita}}, \bibinfo {author} {\bibfnamefont {M.}~\bibnamefont {Abe}},
  \bibinfo {author} {\bibfnamefont {M.}~\bibnamefont {Hada}}, \ and\ \bibinfo
  {author} {\bibfnamefont {Y.}~\bibnamefont {Moriwaki}},\ }\href {\doibase
  10.1088/0953-4075/44/20/209802} {\bibfield  {journal} {\bibinfo  {journal}
  {Journal of Physics B: Atomic, Molecular and Optical Physics}\ }\textbf
  {\bibinfo {volume} {44}},\ \bibinfo {pages} {209802} (\bibinfo {year}
  {2011})}\BibitemShut {NoStop}%
\bibitem [{\citenamefont {Murad}(1982)}]{Murad1982}%
  \BibitemOpen
  \bibfield  {author} {\bibinfo {author} {\bibfnamefont {E.}~\bibnamefont
  {Murad}},\ }\href {\doibase 10.1063/1.444009} {\bibfield  {journal} {\bibinfo
   {journal} {The Journal of Chemical Physics}\ }\textbf {\bibinfo {volume}
  {77}},\ \bibinfo {pages} {2057} (\bibinfo {year} {1982})}\BibitemShut
  {NoStop}%
\bibitem [{\citenamefont {West}()}]{VMI}%
  \BibitemOpen
  \bibfield  {author} {\bibinfo {author} {\bibfnamefont {E.}~\bibnamefont
  {West}},\ }\href@noop {} {}\bibinfo {note} {In preparation}\BibitemShut
  {NoStop}%
\bibitem [{\citenamefont {Mejrissi}\ \emph {et~al.}(2013)\citenamefont
  {Mejrissi}, \citenamefont {Habli}, \citenamefont {Ghalla}, \citenamefont
  {Oujia},\ and\ \citenamefont {Gadéa}}]{BaH+}%
  \BibitemOpen
  \bibfield  {author} {\bibinfo {author} {\bibfnamefont {L.}~\bibnamefont
  {Mejrissi}}, \bibinfo {author} {\bibfnamefont {H.}~\bibnamefont {Habli}},
  \bibinfo {author} {\bibfnamefont {H.}~\bibnamefont {Ghalla}}, \bibinfo
  {author} {\bibfnamefont {B.}~\bibnamefont {Oujia}}, \ and\ \bibinfo {author}
  {\bibfnamefont {F.~X.}\ \bibnamefont {Gadéa}},\ }\href {\doibase
  10.1021/jp4025409} {\bibfield  {journal} {\bibinfo  {journal} {The Journal of
  Physical Chemistry A}\ }\textbf {\bibinfo {volume} {117}},\ \bibinfo {pages}
  {5503} (\bibinfo {year} {2013})},\ \bibinfo {note} {pMID:
  23701525}\BibitemShut {NoStop}%
\bibitem [{\citenamefont {Bartlett}\ \emph {et~al.}(2015)\citenamefont
  {Bartlett}, \citenamefont {VanGundy},\ and\ \citenamefont
  {Heaven}}]{Bartlett2015}%
  \BibitemOpen
  \bibfield  {author} {\bibinfo {author} {\bibfnamefont {J.~H.}\ \bibnamefont
  {Bartlett}}, \bibinfo {author} {\bibfnamefont {R.~A.}\ \bibnamefont
  {VanGundy}}, \ and\ \bibinfo {author} {\bibfnamefont {M.~C.}\ \bibnamefont
  {Heaven}},\ }\href {\doibase 10.1063/1.4927007} {\bibfield  {journal}
  {\bibinfo  {journal} {The Journal of Chemical Physics}\ }\textbf {\bibinfo
  {volume} {143}},\ \bibinfo {pages} {044302} (\bibinfo {year}
  {2015})}\BibitemShut {NoStop}%
\bibitem [{\citenamefont {Woon}\ and\ \citenamefont
  {Dunning}(1993)}]{Woon1993}%
  \BibitemOpen
  \bibfield  {author} {\bibinfo {author} {\bibfnamefont {D.~E.}\ \bibnamefont
  {Woon}}\ and\ \bibinfo {author} {\bibfnamefont {T.~H.}\ \bibnamefont
  {Dunning}},\ }\href {\doibase 10.1063/1.464303} {\bibfield  {journal}
  {\bibinfo  {journal} {The Journal of Chemical Physics}\ }\textbf {\bibinfo
  {volume} {98}},\ \bibinfo {pages} {1358} (\bibinfo {year}
  {1993})}\BibitemShut {NoStop}%
\bibitem [{\citenamefont {Rossa}\ \emph {et~al.}(2012)\citenamefont {Rossa},
  \citenamefont {Cabanillas-Vidosa}, \citenamefont {Pino},\ and\ \citenamefont
  {Ferrero}}]{Rossa2012}%
  \BibitemOpen
  \bibfield  {author} {\bibinfo {author} {\bibfnamefont {M.}~\bibnamefont
  {Rossa}}, \bibinfo {author} {\bibfnamefont {I.}~\bibnamefont
  {Cabanillas-Vidosa}}, \bibinfo {author} {\bibfnamefont {G.~A.}\ \bibnamefont
  {Pino}}, \ and\ \bibinfo {author} {\bibfnamefont {J.~C.}\ \bibnamefont
  {Ferrero}},\ }\href {\doibase 10.1063/1.3682283} {\bibfield  {journal}
  {\bibinfo  {journal} {The Journal of Chemical Physics}\ }\textbf {\bibinfo
  {volume} {136}},\ \bibinfo {pages} {064303} (\bibinfo {year}
  {2012})}\BibitemShut {NoStop}%
\bibitem [{\citenamefont {Bernardini}\ \emph {et~al.}(1998)\citenamefont
  {Bernardini}, \citenamefont {Braccini}, \citenamefont {De~Salvo},
  \citenamefont {Di~Virgilio}, \citenamefont {Gaddi}, \citenamefont {Gennai},
  \citenamefont {Genuini}, \citenamefont {Giazotto}, \citenamefont {Losurdo},
  \citenamefont {Pan}, \citenamefont {Pasqualetti}, \citenamefont {Passuello},
  \citenamefont {Popolizio}, \citenamefont {Raffaelli}, \citenamefont
  {Torelli}, \citenamefont {Zhang}, \citenamefont {Bradaschia}, \citenamefont
  {Del~Fabbro}, \citenamefont {Ferrante}, \citenamefont {Fidecaro},
  \citenamefont {La~Penna}, \citenamefont {Mancini}, \citenamefont {Poggiani},
  \citenamefont {Narducci}, \citenamefont {Solina},\ and\ \citenamefont
  {Valentini}}]{airbake}%
  \BibitemOpen
  \bibfield  {author} {\bibinfo {author} {\bibfnamefont {M.}~\bibnamefont
  {Bernardini}}, \bibinfo {author} {\bibfnamefont {S.}~\bibnamefont
  {Braccini}}, \bibinfo {author} {\bibfnamefont {R.}~\bibnamefont {De~Salvo}},
  \bibinfo {author} {\bibfnamefont {A.}~\bibnamefont {Di~Virgilio}}, \bibinfo
  {author} {\bibfnamefont {A.}~\bibnamefont {Gaddi}}, \bibinfo {author}
  {\bibfnamefont {A.}~\bibnamefont {Gennai}}, \bibinfo {author} {\bibfnamefont
  {G.}~\bibnamefont {Genuini}}, \bibinfo {author} {\bibfnamefont
  {A.}~\bibnamefont {Giazotto}}, \bibinfo {author} {\bibfnamefont
  {G.}~\bibnamefont {Losurdo}}, \bibinfo {author} {\bibfnamefont {H.~B.}\
  \bibnamefont {Pan}}, \bibinfo {author} {\bibfnamefont {A.}~\bibnamefont
  {Pasqualetti}}, \bibinfo {author} {\bibfnamefont {D.}~\bibnamefont
  {Passuello}}, \bibinfo {author} {\bibfnamefont {P.}~\bibnamefont
  {Popolizio}}, \bibinfo {author} {\bibfnamefont {F.}~\bibnamefont
  {Raffaelli}}, \bibinfo {author} {\bibfnamefont {G.}~\bibnamefont {Torelli}},
  \bibinfo {author} {\bibfnamefont {Z.}~\bibnamefont {Zhang}}, \bibinfo
  {author} {\bibfnamefont {C.}~\bibnamefont {Bradaschia}}, \bibinfo {author}
  {\bibfnamefont {R.}~\bibnamefont {Del~Fabbro}}, \bibinfo {author}
  {\bibfnamefont {I.}~\bibnamefont {Ferrante}}, \bibinfo {author}
  {\bibfnamefont {F.}~\bibnamefont {Fidecaro}}, \bibinfo {author}
  {\bibfnamefont {P.}~\bibnamefont {La~Penna}}, \bibinfo {author}
  {\bibfnamefont {S.}~\bibnamefont {Mancini}}, \bibinfo {author} {\bibfnamefont
  {R.}~\bibnamefont {Poggiani}}, \bibinfo {author} {\bibfnamefont
  {P.}~\bibnamefont {Narducci}}, \bibinfo {author} {\bibfnamefont
  {A.}~\bibnamefont {Solina}}, \ and\ \bibinfo {author} {\bibfnamefont
  {R.}~\bibnamefont {Valentini}},\ }\href {\doibase 10.1116/1.580967}
  {\bibfield  {journal} {\bibinfo  {journal} {Journal of Vacuum Science \&
  Technology A}\ }\textbf {\bibinfo {volume} {16}},\ \bibinfo {pages} {188}
  (\bibinfo {year} {1998})}\BibitemShut {NoStop}%
\bibitem [{\citenamefont {Mahner}\ \emph {et~al.}(2003)\citenamefont {Mahner},
  \citenamefont {Hansen}, \citenamefont {Laurent},\ and\ \citenamefont
  {Madsen}}]{PhysRevSTAB.6.013201}%
  \BibitemOpen
  \bibfield  {author} {\bibinfo {author} {\bibfnamefont {E.}~\bibnamefont
  {Mahner}}, \bibinfo {author} {\bibfnamefont {J.}~\bibnamefont {Hansen}},
  \bibinfo {author} {\bibfnamefont {J.-M.}\ \bibnamefont {Laurent}}, \ and\
  \bibinfo {author} {\bibfnamefont {N.}~\bibnamefont {Madsen}},\ }\href
  {\doibase 10.1103/PhysRevSTAB.6.013201} {\bibfield  {journal} {\bibinfo
  {journal} {Phys. Rev. ST Accel. Beams}\ }\textbf {\bibinfo {volume} {6}},\
  \bibinfo {pages} {013201} (\bibinfo {year} {2003})}\BibitemShut {NoStop}%
\bibitem [{\citenamefont {Yang}\ \emph {et~al.}(2020)\citenamefont {Yang},
  \citenamefont {Cheng}, \citenamefont {Peng}, \citenamefont {Wu},
  \citenamefont {Wang}, \citenamefont {Qu},\ and\ \citenamefont {Zhang}}]{BeH}%
  \BibitemOpen
  \bibfield  {author} {\bibinfo {author} {\bibfnamefont {Y.~K.}\ \bibnamefont
  {Yang}}, \bibinfo {author} {\bibfnamefont {Y.}~\bibnamefont {Cheng}},
  \bibinfo {author} {\bibfnamefont {Y.~G.}\ \bibnamefont {Peng}}, \bibinfo
  {author} {\bibfnamefont {Y.}~\bibnamefont {Wu}}, \bibinfo {author}
  {\bibfnamefont {J.~G.}\ \bibnamefont {Wang}}, \bibinfo {author}
  {\bibfnamefont {Y.~Z.}\ \bibnamefont {Qu}}, \ and\ \bibinfo {author}
  {\bibfnamefont {S.~B.}\ \bibnamefont {Zhang}},\ }\href {\doibase
  https://doi.org/10.1016/j.jqsrt.2020.107203} {\bibfield  {journal} {\bibinfo
  {journal} {Journal of Quantitative Spectroscopy and Radiative Transfer}\
  }\textbf {\bibinfo {volume} {254}},\ \bibinfo {pages} {107203} (\bibinfo
  {year} {2020})}\BibitemShut {NoStop}%
\bibitem [{\citenamefont {Ghalila}\ \emph {et~al.}(2008)\citenamefont
  {Ghalila}, \citenamefont {Lahmar}, \citenamefont {Lakhdar},\ and\
  \citenamefont {Hochlaf}}]{Ghalila_2008}%
  \BibitemOpen
  \bibfield  {author} {\bibinfo {author} {\bibfnamefont {H.}~\bibnamefont
  {Ghalila}}, \bibinfo {author} {\bibfnamefont {S.}~\bibnamefont {Lahmar}},
  \bibinfo {author} {\bibfnamefont {Z.~B.}\ \bibnamefont {Lakhdar}}, \ and\
  \bibinfo {author} {\bibfnamefont {M.}~\bibnamefont {Hochlaf}},\ }\href
  {\doibase 10.1088/0953-4075/41/20/205101} {\bibfield  {journal} {\bibinfo
  {journal} {Journal of Physics B: Atomic, Molecular and Optical Physics}\
  }\textbf {\bibinfo {volume} {41}},\ \bibinfo {pages} {205101} (\bibinfo
  {year} {2008})}\BibitemShut {NoStop}%
\bibitem [{\citenamefont {H{\o}jbjerre}\ \emph {et~al.}(2009)\citenamefont
  {H{\o}jbjerre}, \citenamefont {Hansen}, \citenamefont {Skyt}, \citenamefont
  {Staanum},\ and\ \citenamefont {Drewsen}}]{H_jbjerre_2009}%
  \BibitemOpen
  \bibfield  {author} {\bibinfo {author} {\bibfnamefont {K.}~\bibnamefont
  {H{\o}jbjerre}}, \bibinfo {author} {\bibfnamefont {A.~K.}\ \bibnamefont
  {Hansen}}, \bibinfo {author} {\bibfnamefont {P.~S.}\ \bibnamefont {Skyt}},
  \bibinfo {author} {\bibfnamefont {P.~F.}\ \bibnamefont {Staanum}}, \ and\
  \bibinfo {author} {\bibfnamefont {M.}~\bibnamefont {Drewsen}},\ }\href
  {\doibase 10.1088/1367-2630/11/5/055026} {\bibfield  {journal} {\bibinfo
  {journal} {New Journal of Physics}\ }\textbf {\bibinfo {volume} {11}},\
  \bibinfo {pages} {055026} (\bibinfo {year} {2009})}\BibitemShut {NoStop}%
\bibitem [{\citenamefont {Maatouk}\ \emph {et~al.}(2011)\citenamefont
  {Maatouk}, \citenamefont {Houria}, \citenamefont {Yazidi}, \citenamefont
  {Jaidane},\ and\ \citenamefont {Hochlaf}}]{Maatouk_2011}%
  \BibitemOpen
  \bibfield  {author} {\bibinfo {author} {\bibfnamefont {A.}~\bibnamefont
  {Maatouk}}, \bibinfo {author} {\bibfnamefont {A.~B.}\ \bibnamefont {Houria}},
  \bibinfo {author} {\bibfnamefont {O.}~\bibnamefont {Yazidi}}, \bibinfo
  {author} {\bibfnamefont {N.}~\bibnamefont {Jaidane}}, \ and\ \bibinfo
  {author} {\bibfnamefont {M.}~\bibnamefont {Hochlaf}},\ }\href {\doibase
  10.1088/0953-4075/44/22/225101} {\bibfield  {journal} {\bibinfo  {journal}
  {Journal of Physics B: Atomic, Molecular and Optical Physics}\ }\textbf
  {\bibinfo {volume} {44}},\ \bibinfo {pages} {225101} (\bibinfo {year}
  {2011})}\BibitemShut {NoStop}%
\bibitem [{\citenamefont {Kimura}\ \emph {et~al.}(2017)\citenamefont {Kimura},
  \citenamefont {Kajita},\ and\ \citenamefont {Okada}}]{Kimura_2017}%
  \BibitemOpen
  \bibfield  {author} {\bibinfo {author} {\bibfnamefont {N.}~\bibnamefont
  {Kimura}}, \bibinfo {author} {\bibfnamefont {M.}~\bibnamefont {Kajita}}, \
  and\ \bibinfo {author} {\bibfnamefont {K.}~\bibnamefont {Okada}},\ }\href
  {\doibase 10.1088/1742-6596/875/3/022042} {\bibfield  {journal} {\bibinfo
  {journal} {Journal of Physics: Conference Series}\ }\textbf {\bibinfo
  {volume} {875}},\ \bibinfo {pages} {022042} (\bibinfo {year}
  {2017})}\BibitemShut {NoStop}%
\bibitem [{\citenamefont {Rugango}\ \emph {et~al.}(2016)\citenamefont
  {Rugango}, \citenamefont {Calvin}, \citenamefont {Janardan}, \citenamefont
  {Shu},\ and\ \citenamefont {Brown}}]{Rugango2016}%
  \BibitemOpen
  \bibfield  {author} {\bibinfo {author} {\bibfnamefont {R.}~\bibnamefont
  {Rugango}}, \bibinfo {author} {\bibfnamefont {A.~T.}\ \bibnamefont {Calvin}},
  \bibinfo {author} {\bibfnamefont {S.}~\bibnamefont {Janardan}}, \bibinfo
  {author} {\bibfnamefont {G.}~\bibnamefont {Shu}}, \ and\ \bibinfo {author}
  {\bibfnamefont {K.~R.}\ \bibnamefont {Brown}},\ }\href {\doibase
  10.1002/cphc.201600645} {\bibfield  {journal} {\bibinfo  {journal}
  {Chemphyschem : a European journal of chemical physics and physical
  chemistry}\ }\textbf {\bibinfo {volume} {17}},\ \bibinfo {pages}
  {3764—3768} (\bibinfo {year} {2016})}\BibitemShut {NoStop}%
\bibitem [{\citenamefont {Khalil}\ \emph {et~al.}(2013)\citenamefont {Khalil},
  \citenamefont {Le~Quéré}, \citenamefont {Léonard},\ and\ \citenamefont
  {Brites}}]{Khalil2013}%
  \BibitemOpen
  \bibfield  {author} {\bibinfo {author} {\bibfnamefont {H.}~\bibnamefont
  {Khalil}}, \bibinfo {author} {\bibfnamefont {F.}~\bibnamefont {Le~Quéré}},
  \bibinfo {author} {\bibfnamefont {C.}~\bibnamefont {Léonard}}, \ and\
  \bibinfo {author} {\bibfnamefont {V.}~\bibnamefont {Brites}},\ }\href
  {\doibase 10.1021/jp407811c} {\bibfield  {journal} {\bibinfo  {journal} {The
  Journal of Physical Chemistry A}\ }\textbf {\bibinfo {volume} {117}},\
  \bibinfo {pages} {11254} (\bibinfo {year} {2013})}\BibitemShut {NoStop}%
\bibitem [{\citenamefont {{Abu el kher}}\ \emph {et~al.}(2021)\citenamefont
  {{Abu el kher}}, \citenamefont {Zeid}, \citenamefont {El-Kork},\ and\
  \citenamefont {Korek}}]{ABUELKHER2021101264}%
  \BibitemOpen
  \bibfield  {author} {\bibinfo {author} {\bibfnamefont {N.}~\bibnamefont {{Abu
  el kher}}}, \bibinfo {author} {\bibfnamefont {I.}~\bibnamefont {Zeid}},
  \bibinfo {author} {\bibfnamefont {N.}~\bibnamefont {El-Kork}}, \ and\
  \bibinfo {author} {\bibfnamefont {M.}~\bibnamefont {Korek}},\ }\href
  {\doibase https://doi.org/10.1016/j.jocs.2020.101264} {\bibfield  {journal}
  {\bibinfo  {journal} {Journal of Computational Science}\ }\textbf {\bibinfo
  {volume} {51}},\ \bibinfo {pages} {101264} (\bibinfo {year}
  {2021})}\BibitemShut {NoStop}%
\bibitem [{\citenamefont {Sugiyama}\ and\ \citenamefont
  {Yoda}(1997)}]{Sugiyama1997}%
  \BibitemOpen
  \bibfield  {author} {\bibinfo {author} {\bibfnamefont {K.}~\bibnamefont
  {Sugiyama}}\ and\ \bibinfo {author} {\bibfnamefont {J.}~\bibnamefont
  {Yoda}},\ }\href {\doibase 10.1103/PhysRevA.55.R10} {\bibfield  {journal}
  {\bibinfo  {journal} {Phys. Rev. A}\ }\textbf {\bibinfo {volume} {55}},\
  \bibinfo {pages} {R10} (\bibinfo {year} {1997})}\BibitemShut {NoStop}%
\bibitem [{\citenamefont {Hoang}\ \emph {et~al.}(2020)\citenamefont {Hoang},
  \citenamefont {Jau}, \citenamefont {Overstreet},\ and\ \citenamefont
  {Schwindt}}]{Hoang2020}%
  \BibitemOpen
  \bibfield  {author} {\bibinfo {author} {\bibfnamefont {T.~M.}\ \bibnamefont
  {Hoang}}, \bibinfo {author} {\bibfnamefont {Y.-Y.}\ \bibnamefont {Jau}},
  \bibinfo {author} {\bibfnamefont {R.}~\bibnamefont {Overstreet}}, \ and\
  \bibinfo {author} {\bibfnamefont {P.~D.~D.}\ \bibnamefont {Schwindt}},\
  }\href {\doibase 10.1103/PhysRevA.101.022705} {\bibfield  {journal} {\bibinfo
   {journal} {Phys. Rev. A}\ }\textbf {\bibinfo {volume} {101}},\ \bibinfo
  {pages} {022705} (\bibinfo {year} {2020})}\BibitemShut {NoStop}%
\end{thebibliography}%

\end{document}